\documentclass[10pt,journal]{IEEEtran}
\usepackage{bbm}
\usepackage{amsfonts}
\usepackage{tipa}
\usepackage{amssymb}
\usepackage{mathrsfs}
\usepackage{stfloats}
\usepackage{color, soul}
\usepackage[lined,boxed,commentsnumbered]{algorithm2e}
\usepackage{graphicx}
\usepackage{subfigure}
\usepackage{amsfonts,amssymb,amsmath}
\usepackage{latexsym}
\usepackage{multirow}
\usepackage{hyperref}
\usepackage{epsfig,epstopdf}
\usepackage{array}
\newtheorem{remark}{Remark}

\begin{document}
\title{Distortion Minimization for Relay Assisted Wireless Multicast}
%\author{\IEEEauthorblockN{Zhi Chen\IEEEauthorrefmark{1} and Pin-Han Ho\IEEEauthorrefmark{2}}
%\IEEEauthorblockA{Department of Electrical and Computer Engineering\\University of
%Waterloo, Waterloo, Ontario, Canada\\
%Email: \IEEEauthorrefmark{1}z335chen@uwaterloo.ca,
%    \IEEEauthorrefmark{2}\{p4ho\}@engmail.uwaterloo.ca}
%}
\author{Zhi Chen, Pin-Han Ho, and Limei Peng
\thanks{The authors declare that there is no conflict of interest regarding the publication of this paper. Z. Chen and P.-H. Ho are with the Department of Electrical and Computer Engineering,
University of Waterloo, Waterloo, Ontario, N2L 3G1, Canada (e-mail: {z335chen, p4ho}@uwaterloo.ca).

Limei Peng is with Department of Industry Engineering, Ajou University, Korea (email: auroraplm@ajou.ac.kr).}}

\maketitle
\begin{abstract}
%\baselineskip 18pt
%\boldmath
The paper studies the scenario of wireless multicast with a
single transmitter and a relay that deliver scalable source symbols
to the receivers in a decode-and-forward (DF) fashion.
With the end-to-end mean square
error distortion (EED) as performance metric, we firstly derive
the EED expression for the $L$-resolution scalable source symbol
for any receiver. An optimization problem in minimizing the weighted EED is
then formulated for finding the power allocations for all resolution layers at the
transmitter and the relay. Due to nonlinearity of the formulations,
we solve the formulated optimization problems using a
generalized programming algorithm for obtaining good sub-optimal
solutions. Case studies are conducted to verify the proposed
formulations and solution approaches. The results demonstrate
the advantages of the proposed strategies in the relay-assisted
wireless networks for scalable source multicast.
\end{abstract}


\IEEEpeerreviewmaketitle

\section{Introduction}
Wireless communication
suffers from multipath fading and the time-varying characteristic,
which causes distortion to the delivered information.
%This has
%imposed a stringent limitation on some applications where
%retransmission is not possible, such as broadcast/multicast.
To resolve this problem, joint source-channel coding (JSCC) is shown promising in
practical wireless communication systems.
JSCC pairs scalable source coding (SSC) and superposition channel
coding (SPC) by mapping the
source symbols to multiple successively refined channel symbols.
The source from a common source coding block is scalably encoded into several
resolutions/layers and then mapped into successively
refined source symbols at the transmitter.
These source symbols
are then mapped into channel symbols and superimposed
into one JSCC symbol under SPC for
transmission. Employing successive
interference cancelation (SIC), each receiver can then
recover up to a specific layer of JSCC symbols with respect to its channel
quality \cite{Ng-icc}-\cite{Gunduz-tit}.

%Enjoying its multi-resolution in nature, JSCC has been reported for its superb capability
%in mitigating various vicious impacts due to
%fast channel fading and multi-user channel diversity in the
%wireless multicast scenario.
%The multi-granularity decoding process based on successive
%interference cancelation (SIC) allows a receiver of a
%specific channel quality to receive a portion of bits
%in a symbol/data block instead of losing all the information
%according to its instantaneous channel capacity.

In \cite{Ng-icc}, the expected distortion of transmission of a
Gaussian source over a slow fading channel with
only a finite number of fading states is investigated.
In \cite{Tian-tit}, the problem of transmitting a Gaussian
source on a slowly fading Gaussian channel is studied,
subject to the mean squared
error distortion measure.
\cite{Kostina-isit} finds new tight finite block-length bounds
for the best achievable lossy joint source-channel code rate.
In \cite{Gunduz-tit}, reliable transmission of
a discrete memoryless source to multiple destinations
over a relay network is considered, where the relays and
the destinations all have access to side information correlated
with the underlying source signal.

Numerous efforts have been claimed on
multi-resolution SPC over wireless transmissions
\cite{Zhihai-T-IP}-\cite{Ho-twc}.
In \cite{Zhihai-T-IP}, bit allocation for a joint
source/channel video codec over static channels
is studied, and the expected distortion is minimized through
the distribution of the available bits among the subbands.
In \cite{Kondi-T-IP}, the optimal selection
of the JSCC rates through all layers over Rayleigh fading
channels is presented in terms of the overall distortion.
%In \cite{Zhihai-T-csvt}, a rate-distortion (R-D) model for DCT-based video coding is developed and a statistical model is presented to estimate distortion induced by channel errors.
%In \cite{Hossain-tvt}, an adaptive layered modulation scheme for simultaneous voice and data transmission over fading channels is introduced.
%In \cite{Fresnedo-cl}, a two step decoding approach is presented to achieve the near-optimal performance of minimum mean square error (MMSE) decoding under all SNR regimes.
%In \cite{Brante}, spatial diversity was utilized to improve the performance of analog JSCC system in wireless fading channels, where both ML detection and MMSE detection were employed.
%In \cite{Hu-Tcom}, the performance of a discrete-time all-analog-processing JSCC system over AWGN channels is investigated by employing MMSE decoding.
The energy efficiency of various JSCC problems
is studied in \cite{Jain-Tit}. In addition, the JSCC problem of using hybrid digital
analog codes in transmitting a Gaussian source over a Gaussian
channel is studied in \cite{Wilson-Tit}.
\cite{cheng-Wiad} presents a distributed
JSCC system for relay systems
exploiting spatial and temporal correlations.
In \cite{Xiang-tit}, an optimal noise channel quantization
with
random index assignment in a single-layer tandem source channel coding
system with one-level resolution is investigated.
In \cite{Ho-twc}, the SPC transmission
for scalable sources of only two information
layers over relay channels is investigated,
and
the performance improvement is evaluated.
\cite{Ji-Tmultimeida} presents an optimal
JSCC broadcasting scheme by providing different QoS metrics
for heterogeneous users in the network by employing fountain codes.
In \cite{Ji-TSP}, a JSCC model over the MIMO broadcast channel is
presented to minimize the sum mean square error distortion,
where the perfect channel state information
(CSI) is assumed to be available at the transmitter and
at the receivers.
%\cite{Sethakaset-tcom} provides analysis on end-to-end distortion (EED) in an amplify-and-forward multi-relay aided unicast network with two-level resolution sources.
\cite{Kim-ASILOMAR} investigates the
average throughput and distortion of a multi-relay
aided unicast system with two-layer resolution sources,
where the direct link is assumed to be unavailable.

It is clear that most existing works on JSCC have focused on
theoretical performance such as channel capacity,
outage probability and
distortion exponent, which may nonetheless fail
to faithfully reflect the end-to-end (E2E) service quality
in terms of symbol error rates. Further, those theoretical performance metrics are only
suitable as metrics under high signal-to-noise ratio (SNR) regime,
which however are not suitable
under low SNR regime with high channel error probabilities.

%To correctly evaluate the system performance,
%end-to-end distortion (EED) is considered in \cite{Xiang-tit}\cite{Ho-twc}, in which
%both the source quantization error
%as well as channel errors are jointly considered, thus being
%able to effectively reflect the transmission performance in
%arbitrary SNR regimes.
%With such a premise, \cite{Ho-twc} investigates a simple
%two-resolution JSCC relay network consisting of one source, one relay, and
%one destination. It numerically demonstrates the difference between using EED and channel capacity
%as the performance metric, where up to $30\%$ of
%EED reduction is achieved when the optimization process takes
%channel capacity as the metric. Note that the optimization performs
%in \cite{Ho-twc} is mainly for power allocation for two-layer JSCC under
%a fixed relay location, while relay placement with more layers
%has not been explored.

%It is noticeable that the optimization performed in \cite{Ho-twc}
%is mainly for deriving the power allocated for two-layer
%JSCC under a fixed relay location, while relay placement with more layers has not been explored.

In this paper, we are committed in a new research initiative on the
resource allocation problems for multicasting of JSCC symbols,
where the transmit power of each resolution layer at both transmitter and relay are
jointly determined. The aim is to minimize the weighted averaged EED of all users.

The contributions of this work are summarized as follows,
\begin{itemize}
\item introduce a general framework
of JSCC transmission for successively refined information source
with an arbitrary number of layers
over relay networks.
\item formulate an optimization problem for
jointly determining the power assigned to
all resolution layers at both the source node and relay for the weighted averaged EED of all users.
\item
develop a generalized programming algorithm
that employs a Lagrangian dual method to solve
the formulated problems to obtain a good sub-optimal solution.
\end{itemize}

The rest of this paper is organized as follows.
Section II presents the system model.
Section III provides the EED model.
Section IV presents the formulated optimization problems
under various target functions, along with the proposed generalized programming
algorithm to obtain good sub-optimal solutions.
Simulation results are
presented in Section V and the paper is concluded in Section VI.
\begin{figure}[t]
   \centering
   \includegraphics[width = 6.4cm]{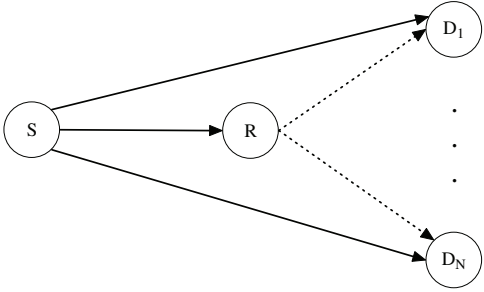}
   \caption{The relayed multicast network considered in the system model,
   where the solid line denotes the transmission in the first time slot
   and the dash line denotes transmission in the second slot.} \label{fig:description}
   \end{figure}

\begin{figure}[t]
   \centering
   \includegraphics[width = 7.2cm]{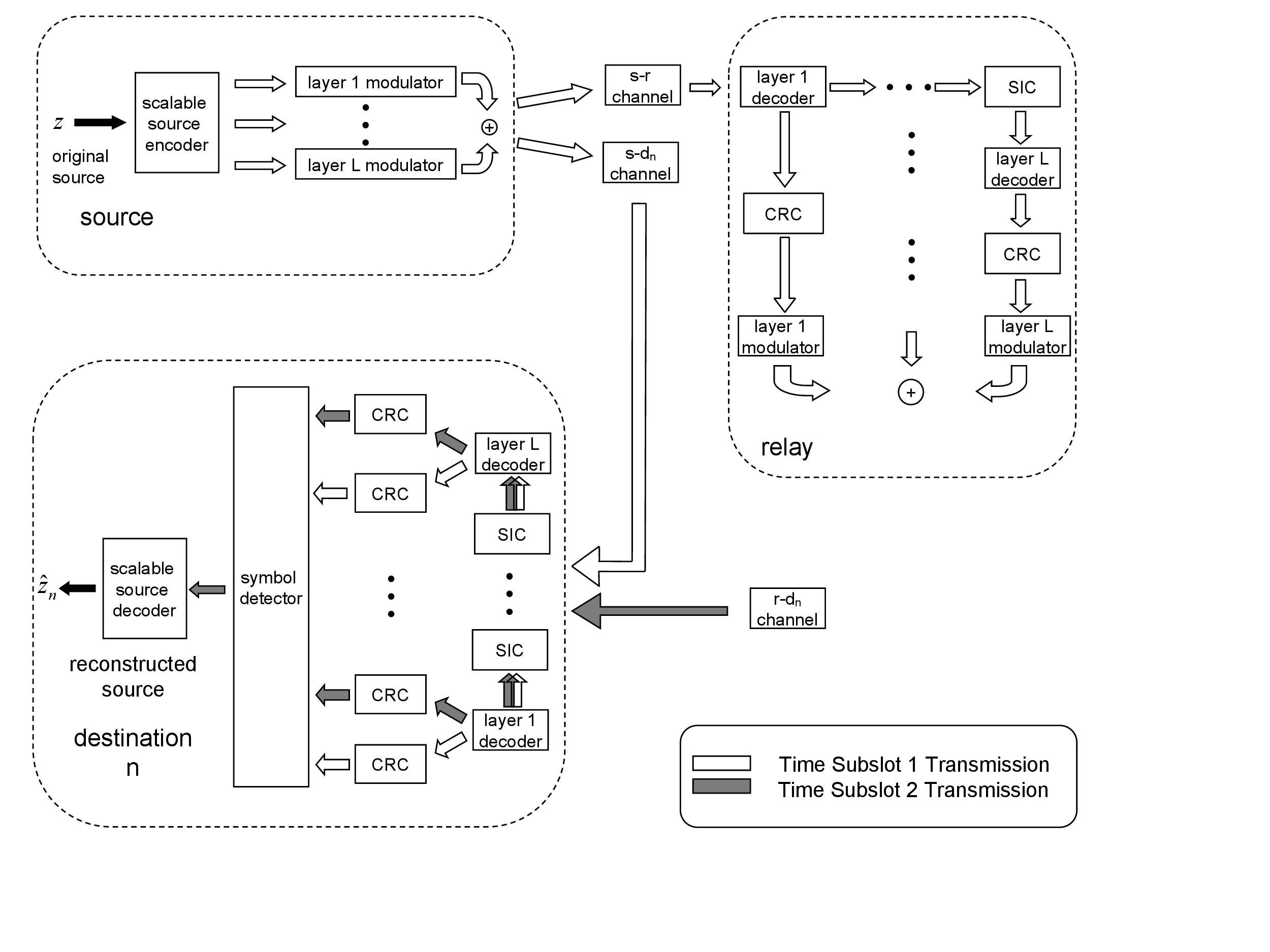}
   \caption{The encoding/decoding structure of scalably encoded
   sources with successive refinement of
   a general $L$-resolution multicast, where the
   received signals from the direct link and the relay links
are decoded
   separately at each destination.
At the scalable source encoder at source node,
the information is encoded at different resolutions/layers
with the associated CRC part attached to the constructed message
in help of decoding at the CRC module of the
relay and destination nodes.
The received messages at each node are then fed into the
cyclic redundance codes (CRC) module for error detection and then
processed at the symbol detector for information reconstruction.
It is also noted that each time slot
is equally divided into two time subslots, where the first subslot is used for source
transmission and the second subslot is used for relay transmission, respectively.} \label{fig:sys}
   \end{figure}

\begin{table*}[ht]
\caption{The notion table of the parameters in this work}
\centering
\small
\begin{tabular}{ | c | c | }
\hline			
  $N$ & the number of receivers/destination nodes \\
\hline	
  $L$ & the number of refined layers for a source symbol \\
  \hline
  $\mathtt{P}_i$ & the average transmit power constraint at node $i$\\
\hline	
  $P_i$ & the transmit power at node $i$  \\
\hline
$\beta_{l}^{(i)}$  & the ratio of power assigned to layer $l$ at node $i$\\
\hline
$\boldsymbol{\beta^{(i)}}$ & the power allocation vector of node $i$\\
\hline
$h_{ij}$ & the channel gain of link $i$-$j$ \\
\hline
$p_{err,d_n}^{(l)}$ & the E2E SER of up to layer $l$ at the end of the second slot
at receiver $d_n$\\
\hline
${\tt p}_{d_n}^{(l)}$ & the realization probability of only to layer $l$ information successfully decoded at $d_n$ \\
\hline
$N_0/2$ & the two-sided power spectral density of the additive white Gaussian noise \\
\hline
$EED_{d_n}$ & the end-to-end distortion at $d_n$ given channel realizations and power allocations. \\
\hline
\end{tabular}
\label{table:notion}
\end{table*}

\section{System Description}
\begin{figure}[t]
   \centering
   \includegraphics[width=7.2cm]{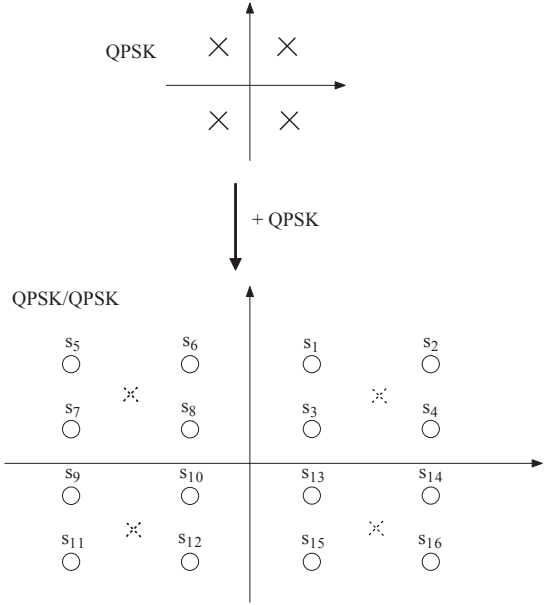}
   \caption{The procedure of SPC constellation with
   two layers using QPSK and QPSK signals as the base
   layer and the enhancement layer, respectively.} \label{fig:three_layer}
   \end{figure}

As shown in Fig. \ref{fig:description}, a relay-aided multicast
network consists of one
source node, one relay and $N$ destination nodes.
The general encoding/decoding structure of a JSCC system
is shown in Fig. \ref{fig:sys}.
The discrete-time, real-valued continuous Gaussian information source
at the source node is scalably
encoded and mapped into successively refined symbols of $L$ layers,
and is broadcasted via both the direct ($s \rightarrow d_n$, $n=1,\ldots,N$) links and the
relayed ($s \rightarrow r$ and $r \rightarrow d_n$) links to enable the multi-resolution information
reconstruction at the destination nodes.
The $L$ layers mapped by the scalable information source are
correspondingly denoted by the base layer, the first enhancement layer,
$\ldots$, and
the ($L-1$)th enhancement layer, respectively. Among them,
the base layer alone provides the lowest resolution
reconstruction. By adding more enhancement layers incorporating with the base layer
gradually improves
the resolution reconstruction to a higher level,
while adding the base layer and all $L-1$
enhancement layers altogether provide
the highest resolution reconstruction.
Before transmission, each source symbol corresponding to each layer is mapped into
one or a set of channel symbols and the channel symbols of all the
symbols of $L$ layers are further superimposed under SPC.
For reference, a typical example showing
the procedure of SPC constellation of the symbols of each layer for a
two-layer case
is shown in Fig. \ref{fig:three_layer}.
Further, for readability, the definitions of some parameters used in this work are listed in Table. \ref{table:notion}.

Let
the average transmit power constraint at the source node and the relay node
be denoted by $\mathtt{P}_s$ and $\mathtt{P}_r$.
In addition, the
actual transmit power at source and relay are denoted by $P_s$ and $P_r$,
respectively.
For the power assigned to each layer,
it is assumed that $\beta_{1}^{(i)}P_i$ ($i=s,r$)
is allocated to the base layer at node $i$, and
$\beta_{l}^{(i)}P_i$ ($l=2,\ldots,L$) is allocated to the $(l-1)$th
enhancement layer, where
$\beta_{L}^{(i)}=1-\sum_{l=1}^{L-1}\beta_{l}^{(i)} \ge 0$ is for
the $(L-1)$th enhancement layer and it is physically
required that $\beta_{l}^{(i)} \ge 0$ ($\forall i,l$). %Note also that
The power allocation vector at node $i$ is
also referred to as
$\boldsymbol{\beta^{(i)}}=\{ \beta_{l}^{(i)}|l=1,\cdots,L \}$
for simplicity.
Note that
both the
vector representation and the scalar representation of
the power assignment parameter
will be used interchangeably in the following
analysis.

The general encoding/decoding structure of a JSCC system is shown
in Fig. \ref{fig:sys}.
In the first time subslot, the source node broadcasts
the JSCC symbols to the relay node and all the receivers.
In the second time subslot, the source node keeps silent and
the relay node
broadcasts its decoded and re-encoded SPC symbols to all destination nodes,
possibly with a different power allocation ($\beta_{l}^{(r)}$) for
different layer symbols.

When a JSCC symbol is received
(either at the relay or the destination), it is firstly decoded as
if it just contains the base layer symbol while taking all the other
layer energy as interference. Once the base layer is correctly
obtained, it is subtracted from the JSCC symbol, resulting in a new
JSCC symbol, which is in turn decoded as if it is just for the first
enhancement layer symbol while taking all the other layer energy as
interference. Such an iterative and successive decoding process, also
known as successive interference cancelation (SIC), proceeds up to the
$L'$-th layer, which is defined according to the instantaneous channel
quality perceived at the receiver.

Further, it should be noted that,
at the end of the first time subslot,
the received messages of each layer at
relay from the source transmitter is fed into the cyclic redundancy code (CRC) module for error detection
and messages of the successfully decoded layers are re-encoded and
superimposed for transmission to the destinations
at the second subslot. It is also noted that
CRC is performed at the layer level for higher system performance.

%For the decoding at both the relay node and
%the destination nodes, SIC is employed
%to successively decode symbols of each layer.
%The procedure is as follows.
%The receiver firstly
%decodes the JSCC symbol as if it just contains the base layer symbol
%while taking all the other layer energy as interference. Once the
%base layer is correctly obtained, it is subtracted from the JSCC
%symbol, resulting in a new JSCC symbol, which is decoded as if it
%is just for the first enhancement layer symbol while taking all the
%other layer energy as interference. Such an iterative and successive
%decoding process will proceed until all the layered symbols are
%obtained, or when the decoded result in the iteration is found
%incorrect. In addition, it is noted that in this work
%the direct version from the source node and the
%relayed version from the relay node are decoded
%separately at the destination nodes.
%Hence, a destination node successfully decodes
%the $l$ layer as long as it can obtain it from
%either the direct version or the relayed version via SIC.
%Otherwise, the destination node fails in decoding the $l$th layer
%as the $l$th layer in both versions are lost via SIC decoding.

At the end of the second time subslot,
the received information from both the direct
and relay links, respectively, is fed into the cyclic redundancy code (CRC) module
for error detection and then processed by the symbol
selector to determine the reconstructed bits of the
channel use.

With the above mentioned SIC decoding process, a
number of $L+1$ events with different probabilities
can be defined according to the instantaneous channel
capacity, where the first event denotes the loss of all
information layers, the second event denotes only
the base layer is successfully obtained,
and the $l$th ($2 < l \le L+1$) event
denotes that the base layer and enhancement layers
$1,\ldots,l-2$ are successfully obtained.
%At the end of the second time slot,
%the decoded layered information of each
%transmission
%is fed into the CRC module for error
%detection and then processed by the symbol selector to obtain a reconstructed
%version of the original information source. Based on the
%amount of successfully
%decoded layers, a number of $L+1$
%events with different probabilities can be defined,
%with the first event denoting that all information layers are lost, and
%the $j$th ($1 <j \ge L+1$) event denoting that
%the base layer and enhancement layers $1,\ldots,j-1$ are
%successfully decoded.

%Note also that in this work we assume that
%the decoding procedure of the source conveyed version and
%the relayed version at each destination node are conducted
%independently, i.e., the two version are not combined before decoding.
%Hence the $l$th layer is successfully decoded at
%a destination node
%if it is decoded successfully at
%least in either of the two versions.

%It is also noted that only the statistical channel state
%information of
%all links is assumed to be available at the source node in this work,
%which enables it to
%derive
%the appropriate power allocations to each layer at itself and the
%relay node.
%Some feedback and signaling are therefore required, as the source node
%needs feedback of the statistical CSI information of all links from the
%destination nodes and the relay node,
%and then
%notify the relay node how to optimally allocate its power
%to each layered
%information.
%In addition, due to the difficulty to perfectly synchronizes the
%time slots, some overhead to index the source symbols are requested
%to make the reconstructed source messages in order.

\section{System Model}
In this section, we provide an end-to-end distortion (EED) model for an $L$-layer JSCC relay-aided multicast network.

\subsection{Channel Model} \label{sec:channel_model}
Consider a Nakagami-$\rho$ fading channel where $m$
is the Nakagami parameter
representing the severity of the channel fading fluctuations, which
degrades to the special
Rayleigh fading model with $\rho=1$. Further,
it is assumed that he channel gains of all links
remain unchanged within one slot (where one slot consists of
two equally divided sub-slots
for the direct transmission and relayed transmission, respectively)
and are independent from each other
in different slots for different links. In addition, it is
assumed that each transmitter only has the statistics of the channel state
information (CSI).
%When $m > 1$, the Nakagami-$m$
%channel represents a channel that is not severely faded
%compared with Rayleigh fading channels.
%Specifically, with $m \rightarrow \infty$, the
%Nakagami-$m$ channel degrades to a static channel with constant
%gain.

The probability density function (pdf) $f_h(h_{ij})$
and the cumulative density function (cdf) $F_h(h_{ij})$
of the channel power gain over link $i \rightarrow j$ ($i=s,r$ and $j=r,d_1,\ldots,d_N$), are given by,
\begin{align}
&f_h(h_{ij})=(\frac{\rho}{\bar{h}_{ij}})^{\rho}\frac{h_{ij}^{\rho-1}}{\Gamma(\rho)}\exp(-\frac{\rho}{\bar{h}_{ij}}h_{ij}) \\
&F_h(h_{ij})=\frac{\gamma(\rho,\frac{\rho}{\bar{h}_{ij}}h_{ij})}{\Gamma(\rho)}=\frac{1-\Gamma(\rho,\frac{\rho}{\bar{h}_{ij}}h_{ij})}{\Gamma(\rho)}
\end{align}
where $\gamma(\cdot,\cdot)$ and $\Gamma(\cdot,\cdot)$ are the lower
and upper incomplete Gamma functions, respectively.
$h_{ij}$ and $\bar{h}_{ij}$
are the instantaneous channel power gain and its average value,
respectively. In addition, we have $\bar{h}_{ij}=1/d_{ij}^{\alpha}$
accounting for the large-scale fading,
where $\alpha$ is the pass-loss exponent and $d_{ij}$
is the distance between node $i$ and $j$.
Hence the average receiver-side power level is
$\bar{\gamma}_{ij}=P_i\bar{h}_{ij}$ where $P_i$ ($i=s,r$) is the
transmit power at the transmitter $i$.
In addition, we denote $N_0/2$ as the two-sided power spectral density
of the additive white Gaussian noise (AWGN).

In the following, with the assumption that
the power allocated to each layer is specified and the channel gains are known,
the E2E SER of up to layer $l$ at the end of the second slot
at $d_n$, denoted by $p_{err,d_n}^{(l)}$, is hence given by,
\begin{align}
p_{err,d_n}^{(l)}=&p_{sd_n}^{(l)}
\left( 1- \left( 1- p_{sr}^{(l)} \right)
\left( 1- p_{rd_n}^{(l)}  \right) \right), \label{eq:E2E_err_pro}
\end{align}
where $p_{ij}^{(l)}$ is the SER of the $l$th layer infomation over link $i$-$j$.
Note that the product in (\ref{eq:E2E_err_pro})
follows from the independence of transmission of different links
and the terms in the outer bracket denotes the SER probability that
at least one of the $s \rightarrow r$ and $r \rightarrow d_n$ links
fails to provide decodable version of up to layer $l$
in relayed transmission given the realized channels.
For reference, the detailed derivation of E2E SER of (\ref{eq:E2E_err_pro}) as well as that of $p_{ij}^{(l)}$
is presented in Appendix \ref{appendix:1} and are omitted here for brevity.

\subsection{Proposed EED model}
With the allocated power for each layer at the
source and relay as well as the instantaneous channel qualities over link $i \rightarrow j$,
the reconstruction
quality ($RQ$) at receiver $j$ for this
transmission can be divided into $L+1$ categories (referring to decoding quality of the
$L$ layer symbol), namely
${\tt L}_{ij}^{(0)}$, ${\tt L}_{ij}^{(1)}$, $\ldots$ and
${\tt L}_{ij}^{(L)}$,
where ${\tt L}_{ij}^{(L)}$  represents the case that
all layer information are corrected decoded and the symbol is
perfectly reconstructed at node $j$.
${\tt L}_{ij}^{(0)}$ defines the case that all layer information is lost
in this transmission over link $i \rightarrow j$.
In addition, ${\tt L}_{ij}^{(l)}$ ($0< l \le L$)
indicates the case that only the lower $l$ layers are decoded for this transmission, including the base layer
and the $l-1$ lower enhancement layers.
%As defined above,
%the probability that the event ${\tt L}_{ij}^{(l)}$ occurs, i.e.,
%the instantaneous SNR is in category $l$, is denoted by
%$p_{ij}^{(l)}$.

Taking both the direct and
relay links for a specific destination node, i.e., $d_n$ into account,
${\tt L}_{d_n}^{(l)}$ is defined as the category in which
only $l$ lower layers can be successfully decoded
at $d_n$.
Clearly, in category ${\tt L}_{d_n}^{(0)}$, the $s \rightarrow d_n$ link and at
least one of
$s \rightarrow r$ link and $r \rightarrow d_n$ link
are not able to support even the delivery of the base layer
information and the source information is totally lost
at destination node $d_n$.
In category ${\tt L}_{d_n}^{(l)}$ ($0<l<L$), only the base layer and up to the $(l-1)$th enhancement
layer can be successfully obtained
at the destination node $d_n$ through either the direct or the relayed links.
In category ${\tt L}_{d_n}^{(L)}$, all layered source
symbols are successfully reconstructed at
the destination node $d_n$. Given the channel realizations, the associated
realization probabilities
of such events at destination node $d_n$ are
therefore given by,
\begin{align}
&{\tt p}_{d_n}^{(0)}=p_{err,d_n}^{(1)} \label{eq:class_0}\\
&{\tt p}_{d_n}^{(l)}=p_{err,d_n}^{(l+1)}-p_{err,d_n}^{(l)}, \quad l=1,\ldots,L-1 \label{eq:class_j}\\
&{\tt p}_{d_n}^{(L)}=1-p_{err,d_n}^{(L)}
\label{eq:class_L}
\end{align}
where in (\ref{eq:class_j}) the difference between the E2E SER of up to layer $l+1$ and $l$
is the realization probability of successfully decoding only  up to layer $l$.
%where the realization probability of is naturally given by
%the difference between the E2E SER of up to layer $l$

%In addition, taking the distortion of each RQ over the associated link
%into account, the expected EED at the
%$n$th destination node is therefore given by,
%\begin{align}
%D(n) = \sum_{l=0}^{L}\sum_{k=0}^L\sum_{m=0}^L p_{sr}^{(l)}p_{sn}^{(k)}p_{rn}^{(m)}
%D_{lkm}(n) \label{eq:expected EED}
%\end{align}
%where $D_{lkm}(n)$ is the EED for a given RQ group ($l$, $k$, $m$),
%denoting
%that the instantaneous SNRs of the associated links $s \rightarrow r$,$s \rightarrow d_n$
%and $r \rightarrow d_n$ are in SNR realization category $l$, $k$ and $m$ respectively.

%Note that the possible highest layer successfully decoded
%for a specific RQ group ($l$, $k$, $m$) is
%$\max(k,\min(l,m))$,
%as in the relayed transmission ($s \rightarrow r$ and $r \rightarrow d_n$), the destination
%can only decode as many layers as those correctly decoded at the relay node.
%%i.e, $l \ge m$ is physically
%%required.
%Taking both the RQ group
%and the E2E RQ category into account,
%the E2E RQ category ${\tt L}_q$ can also be denoted by the set $A_q$, where
%$A_q=\{(l,k,m)|\max(k,\min(l,m))=q\}$.

\subsection{EED Evaluation} \label{sec:eed}
In \cite{Xiang-tit} and \cite{Ho-twc}, the EED expressions for
one-level and two-level resolution cases are derived, respectively.
The non-asymptotic EED expression
for the realized channel gains of all links
under generally $L$ layers
can be derived in a similar way, which is given in Appendix \ref{appendix:1}.
The associated EED expression given the instantaneous channel gains
that can reconstruct the
information source at destination node $d_n$, denoted by $EED_{d_n}$,
%the reconstruction level up to layer $j$ ($j=0,1,\ldots,L$),
can hence be given by
\begin{align}
%D_{{\tt L}_0}=& \sigma^2 \hat{p}_{err,{\tt L}_1} \label{eq:EED_L_0} \\
&EED_{d_n}(h_{sd_n},h_{sr},h_{rd_n})   \nonumber\\
=& \sum_{l=1}^{L}D_{ Q_l}{\tt p}_{d_n}^{(l)}
+\sigma^2{\tt p}_{d_n}^{(0)} \nonumber \\
=&D_{{ Q}_L}\left(1-p_{err,d_n}^{(L)} \right)
+\sum_{l=1}^{L-1}D_{ Q_l}\left(p_{err,d_n}^{(l+1)}
-p_{err,d_n}^{(l)} \right) \nonumber\\
& \,\,\, + \sigma^2 p_{err,d_n}^{(1)} \label{eq:EED_L_j}
\end{align}
where $\sigma^2$ in this case is the variance of the Gaussian
source signal and
$D_{Q_l}$ denotes the quantization
distortion in the reconstruction of
the scalable source up to layer $l$.
%and $\hat{p}_{err, {\tt L}_j}$ denotes the
%error probability of decoding up to layer $j$.

By considering
a real-valued Gaussian source with unit-variance,
its distortion exponent is denoted by the R-D function $D_{RQ}=2^{-2R}$, where
$R$ is the number of bits of each symbol. Therefore,
combining it with (\ref{eq:EED_L_j}) together
leads to (\ref{eq:eed_all}) as follows,
\begin{align}
&EED_{d_n}(h_{sd_n},h_{sr},h_{rd_n}) \nonumber\\
= &
\sum_{l=1}^{L}2^{-2\sum_{j=1}^{l} R_j}{\tt p}_{d_n}^{(l)}
+\sigma^2{\tt p}_{d_n}^{(0)} \label{eq:eed_all} \\
=& 2^{-2\sum_{j=1}^{L} R_j}\left(1
-p_{err,d_n}^{(L)} \right)  \nonumber\\
&\,\,+ \sum_{l=1}^{L-1}2^{-2\sum_{j=1}^{l} R_j}\left(p_{err,d_n}^{(l+1)}
-p_{err,d_n}^{(l)} \right)
+\sigma^2{\tt p}_{d_n}^{(0)} \nonumber
\end{align}
where $R_l$ is the number of bits allocated to
the $i$th layer per symbol under the $L$-layer
JSCC architecture. For instance, if the base layer employs BPSK,
we have
$R_1=1$ bit for each symbol, i.e.,
the base layer contains $1$ bit information per symbol superimposed with other
upper layer symbols.
%Note that in (\ref{eq:err_distortion_eed}),
%the E2E error probability of up to layer $j$ can be decoded at $d_n$,
%i.e.,
%$\hat{p}_{err,{\tt L}_j}(n)$,
%is given by,
%\begin{align}
%\hat{p}_{err,{\tt L}_j}(n)=&p_{err,sn}^{(j)}
%\left( 1- \left( 1- p_{err,sr}^{(j)} \right)
%\left( 1- p_{err,rn}^{(j)}  \right) \right), \label{eq:end_to_end_err_pro}
%\end{align}
%where $p_{err,tr}^{(j)}$ is the symbol error probability of decoding
%up to layer $j$ at node $k$ from the transmission over link $t \rightarrow r$.
%Note also that for a given RQ group ($l$, $k$, $m$), we have
%\begin{align}
%&p_{err,sr}^{(j)}\approx 1, \quad \mbox{if $l<j$} \\
%&p_{err,sn}^{(j)} \approx 1, \quad \mbox{if $k<j$} \\
%&p_{err,rn}^{(j)} \approx 1, \quad \mbox{if $m<j$}
%%\%end{array} \right.
%\end{align}
%as the probability of successfully decoding up to layer $j$ over
%a link with its
%receiver-side SNR falling into a category whose index is
%less than $j$, is less than the predefined threshold $\epsilon_{th}$
%and is hence negligible. Therefore (\ref{eq:end_to_end_err_pro})
%can be further simplified, if the realized RQs over the associated links
%are in
%lower RQ categories, thus reducing the overall computational efforts.
%\begin{figure*}
%\begin{align}
%D_{lkm}(n)=
%&2^{-2(\sum_{i=1}^{\max(k,\min(l,m))} R_i)}\left(1-\hat{p}_{err,{\tt L}_{\max(k,\min(l,m))}}(n)\right) \nonumber \\
%&+\sum_{j=1}^{\max(k,\min(l,m))-1}2^{-2(\sum_{i=1}^{j} R_i)}
%\left( \hat{p}_{err,{\tt L}_{j+1}}(n)-\hat{p}_{err,{\tt L}_j}(n) \right)
%+\sigma^2 \hat{p}_{err,{\tt L}_1}(n)   %\quad k \ge l>m
%\label{eq:err_distortion_eed}
%\end{align}
%\end{figure*}

By averaging the EED expressions over the
channel realizations of all associated links (S-R,S-$d_n$,R-$d_n$),
%RQ groups ($l,k,m$),
the expected EED at
destination node $d_n$
is given by,
\begin{align}
\overline{EED}_{d_n} = & \iiint\limits_{h_{sd_n},h_{sr},h_{rd_n}}\,
 EED_{d_n}(h_{sd_n},h_{sr},h_{rd_n})
f_h(h_{sd_n}) \nonumber\\
& \,\,f_h(h_{rd_n})f_h(h_{sr})\,
\mathrm{d} h_{sd_n}\, \mathrm{d} h_{sr}\, \mathrm{d} h_{rd_n} \label{eq:eed_expected}
\end{align}

Interestingly, an inequality can be derived from the
EED expressions in (\ref{eq:eed_all}) and (\ref{eq:eed_expected})
as follows.
\begin{align}
&0 < 2^{-2\sum_{j=1}^{L} R_j} \le EED_{d_n}
\le \sigma^2, \quad n=1,\cdots, N
\label{eq:eed_nnequality} \\
&0 < 2^{-2\sum_{j=1}^{L} R_j} \le \overline{EED}_{d_n}
\le \sigma^2, \quad n=1,\cdots, N
\label{eq:eed_nnequality_avg}
\end{align}
where the lower bound holds true when
all $L$ layers are successfully decoded
and the upper bound holds true when all $L$ layers are lost.

\section{Formulation for EED Minimization}
We take the target function
as for minimizing the weighted sum of EEDs of all source-destination pairs.
The optimization problem, termed as {\bf Popt},
is then formulated as follows.
\begin{align}
\min_{\boldsymbol{\beta^{(i)}},P_i} \quad
\sum_{n=1}^{N}c_n\overline{EED}_{d_n} \label{eq:opt_total_EED_prob}
\end{align}
subject to
\begin{align}
&\boldsymbol{\beta^{(i)}} \ge 0, \quad i=s,r \label{eq:opt_constraint_1}\\
&\mathbf{1}^T \boldsymbol{\beta^{(i)}} \le 1, \quad i=s,r  \label{eq:opt_constraint_ratio}\\
&0 \le P_i \le \mathtt{P}_i, \quad i=s,r \label{eq:opt_constraint_2}%\\
\end{align}
where in (\ref{eq:opt_total_EED_prob}) the averaged EED of each user is from
(\ref{eq:eed_expected}) by taking the expected value of EED in (\ref{eq:eed_all}) over all realizations of the associated links. $c_n$ is the predefined weight for the $n$th user and we have $0 \le c_n \le 1$ and $\sum c_n=1$.
(\ref{eq:opt_constraint_1}) gives
the natural non-negative property of
the feasible power allocation vectors and
(\ref{eq:opt_constraint_ratio}) serves as the normalization constraint for the power allocation vectors.
(\ref{eq:opt_constraint_2}) gives
the average transmit power constraint,
where $P_i$ is the transmit power at node $i$ ($i=s,r$)
and should be constrained.
It can be readily found that the optimal solution is achieved when
$P_i=\mathtt{P}_i$ and we can simply replace $P_i$
with $\mathtt{P}_i$ under a numerical method.

Note that {\bf Popt} is a non-convex optimization problem and
the global optimal solution is difficult to be solved.
Henceforth, the Lagrangian dual method is employed to
find the solution to the dual problem of {\bf Popt}, which serves as
a very good lower bound to {\bf Popt}.

Let $\nu^{(i)}$ ($i=s,r$) be the dual variables of {\bf Popt} associated
with the physical normalization constraint
of the power allocation vectors in (\ref{eq:opt_constraint_ratio}),
respectively.
The Lagrangian of problem {\bf P1} can then be expressed as,
\begin{align}
L^{\bf P1}(\boldsymbol{\beta^{(i)}},\nu^{(i)})=
\sum_{n=1}^{N}c_n\overline{EED}_{d_n}
%- \sum_{i=s,r}(\boldsymbol{\mu^{(i)}})^T\boldsymbol{\beta^{(i)}}
+ \sum_{i=s,r}\nu^{(i)}\left(\mathbf{1}^T \boldsymbol{\beta^{(i)}} - 1 \right)
\label{eq:Lagrangian}
\end{align}
subject to $\boldsymbol{\beta^{(i)}} > \boldsymbol{0}$. The associated Lagrangian dual function of
$L^{\bf P1}(\boldsymbol{\beta^{(i)}},\nu^{(i)})$ in (\ref{eq:Lagrangian})
is defined as,
\begin{align}
g^{\bf P1}(\nu^{(i)}) &=
\min_{\boldsymbol{\beta^{(i)}}>\boldsymbol{0}} \quad L^{\bf P1}(\boldsymbol{\beta^{(i)}},\nu^{(i)}). \label{eq:lagrangian_dual}\\
&=\min_{\boldsymbol{\beta^{(i)}}>\boldsymbol{0}} \quad
\sum_{n=1}^{N}c_n\overline{EED}_{d_n}
%- \sum_{i=s,r}(\boldsymbol{\mu^{(i)}})^T\boldsymbol{\beta^{(i)}}
+ \sum_{i=s,r}\nu^{(i)}\left(\mathbf{1}^T \boldsymbol{\beta^{(i)}} - 1 \right) \nonumber
\end{align}

Correspondingly, the Lagrangian dual problem, denoted by {\bf P1-D}, is defined as
\begin{align}
&\max_{\nu^{(i)} \ge 0} \quad
g^{\bf P1}\left(\nu^{(i)}\right) \label{eq:dual_problem}\\
=&\max_{\nu^{(i)} \ge 0 } \quad
\left(\min_{\boldsymbol{\beta^{(i)}}>\boldsymbol{0}} \quad
\sum_{n=1}^{N}c_n\overline{EED}_{d_n}
%- \sum_{i=s,r}(\boldsymbol{\mu^{(i)}})^T\boldsymbol{\beta^{(i)}}
+ \sum_{i=s,r}\nu^{(i)}\left(\mathbf{1}^T \boldsymbol{\beta^{(i)}} - 1
\right) \right)
\nonumber
\end{align}

Let $p^*$ and $d^*$ be the optimal solutions to
the primal problem {\bf Popt} and the associated dual problem {\bf Popt-D}.
According to the weak duality property in \cite{Boyd04}, we have
$p^* \ge d^*$, i.e., $d^*$ serves as a good lower bound to $p^*$ of {\bf Popt}.
In fact, $d^*$ is equal to the optimal solution of the
convexified primal problem of {\bf P1} , i.e., $\hat{p}^*$, which
is defined as follows \cite{Freund04},
\begin{align}
\hat{p}^* = \min\{ p: (0,p) \in C  \}
\end{align}
where $C$ is the convex hull of the feasible region $I$, defined as,
\begin{align}
I=&\{ (s,z)| \,\,\, \exists \boldsymbol{\beta^{(i)}} \ge \boldsymbol{0}
\mbox{ for which }
s \ge \left( \mathbf{1}^T \boldsymbol{\beta^{(i)}} - 1 \right) \nonumber\\
& \mbox{ and }
z \ge \sum_n c_nEED_{d_n}(\boldsymbol{\beta^{(i)}}) \}.
\end{align}
Note that any pair $(0, p) \in I$ is a feasible solution point
for {\bf Popt} and any pair $(0, p) \in C$ is a feasible point
for the convexified primal problem.

To solve the Lagrangian dual problem {\bf Popt-D},
the generalized programming algorithm in \cite{Freund04} by Freund
is employed and is presented as follows.
\begin{enumerate}
\item Initialization:
$E_k=\{\boldsymbol{\beta}_{1}^{(\boldsymbol{i})}, \cdots, \boldsymbol{\beta}_{k}^{(\boldsymbol{i})}\}$ ($i=s,r$), $d_{\min} = -\infty$, $d_{\max}=\infty$.
\item Solve the following linear program below for $\lambda_l$ in the $k$th iteration,
\begin{align}
(\mbox{LP}^k)\quad z_k = \min_{\lambda_l} &\quad \sum_{l=1}^{k} \lambda_l^{(k)} \sum_{n=1}^{N}
c_n\overline{EED}_{d_n}(\boldsymbol{\beta}_{l}^{(\boldsymbol{i})}) \label{eq:generalized_programming_LP_obj} \\
\mbox{s.t.} & \quad \sum_{l=1}^{k} \lambda_l
\left(\mathbf{1}^T \boldsymbol{\beta}_{l}^{(\boldsymbol{i})} - 1 \right) \le 0   \label{eq:generalized_programming_LP_con_1} \\
& \quad \sum_{l=1}^{k} \lambda_l^{(k)} = 1 \label{eq:generalized_programming_LP_con_2} \\
& \quad \lambda_l^{(k)} \ge 0 \label{eq:generalized_programming_LP_con_3}
\end{align}
where we define $s_k = \sum_{l=1}^{k} \lambda_l^{(k)}
\left(\mathbf{1}^T \boldsymbol{\beta}_{l}^{(\boldsymbol{i})} - 1 \right)$
and $\boldsymbol{\lambda}^{(k)}=\{\lambda_l^{(k)}|l=1,\cdots,k\}$.
In addition, we solve
the corresponding dual of the linear programming for
$\nu_k^{(i)}$ and $\Theta_k$,
\begin{align}
(\mbox{DLP}^k) \quad \max_{\nu^{(i)}_k,\Theta_k} & \quad \Theta_k \label{eq:generalized_programming_LP_dual_obj}
\end{align}
subject to
\begin{align}
& \quad \Theta_k \le \sum_{n=1}^{N} c_n\overline{EED}_{d_n}(\boldsymbol{\beta}_{l}^{(\boldsymbol{i})})
+ \sum_{i=s,r}\nu^{(i)}_k\left(\mathbf{1}^T \boldsymbol{\beta}_l^{(\boldsymbol{i})} - 1 \right) \nonumber \\
&\quad\quad\quad\quad l=1,\cdots,k.
  \label{eq:generalized_programming_LP_dual_con1} \\
      & \quad \nu^{(i)}_k \ge 0 \quad i=s,r.\label{eq:generalized_programming_LP_dual_con2}
\end{align}
which can be further re-written as
\begin{align}
(\mbox{DLP}^k) \quad \max_{\nu^{(i)}_k} \quad \min_{\boldsymbol{\beta}_l^{(\boldsymbol{i})}\in E_k}
& \sum_{n=1}^{N} c_n\overline{EED}_{d_n}(\boldsymbol{\beta}_{l}^{(\boldsymbol{i})})  \\
&+ \sum_{i=s,r}\nu^{(i)}_k\left(\mathbf{1}^T \boldsymbol{\beta}_l^{(\boldsymbol{i})} - 1 \right). \nonumber
\end{align}
\item Solve the dual function below for
$\boldsymbol{\beta}_{k+1}^{(\boldsymbol{i})}$ ($i=s,r$),
\begin{align}
g^{\bf P1}(\nu^{(i)}_k)=
\min_{\boldsymbol{\beta}_{k+1}^{(\boldsymbol{i})}>\boldsymbol{0}}\,\,\, &\sum_{n=1}^{N}
c_n\overline{EED}_{d_n}(\boldsymbol{\beta}_{k+1}^{(\boldsymbol{i})}) + \nonumber\\
%- \sum_{i=s,r}(\boldsymbol{\mu^{(i)}})^T\boldsymbol{\beta^{(i)}}
& \sum_{i=s,r}\nu^{(i)}_k\left(\mathbf{1}^T \boldsymbol{\beta}_{k+1}^{(\boldsymbol{i})} - 1 \right). \label{eq:generalized_programming_dual}
\end{align}
\item Shrink the gap between the upper bound $d_{\max}$ and the lower bound $d_{\min}$ by setting $d_{\min} \leftarrow \max\{d_{\min}, g^{\bf P1}(\nu^{(i)}_k) \}$ and
$d_{\max} \leftarrow \min  \{d_{\max}, z_k \}$.
If $d_{\max}-d_{\min} \le \epsilon$ (predefined threshold), go to Step 5,
otherwise we set $E_k=\{\boldsymbol{\beta}_{1}^{(\boldsymbol{i})}, \cdots, \boldsymbol{\beta}_{k}^{(\boldsymbol{i})}, \boldsymbol{\beta}_{k+1}^{(\boldsymbol{i})}\}$
and go to Step 2).
\item Output: $g^{\bf P1}(\nu^{(i)}_k) \}$ and
$\boldsymbol{\beta}_{k+1}^{(\boldsymbol{i})}$ (satisfying the stopping rule in Step 4).
\end{enumerate}
Note that in the initialization step (Step 1)) the $2k$ vectors
$\boldsymbol{\beta}_{j}^{(\boldsymbol{i})}$
($j=1,\cdots,k$ and $i=s,r$) can be arbitrarily
selected as long as the constraints
in (\ref{eq:opt_constraint_1}) and (\ref{eq:opt_constraint_ratio})
are satisfied.
In Step 2) the optimal solutions to the linear programming
and
its associated dual in each iteration are always feasible, i.e.,
there is always a pair ($z_k$, $\lambda_l$) and a pair
($\Theta_k$, $\nu_k^{(i)}$) for ($\mbox{DLP}^k$)
given $\nu_k^{(i)} \ge 0$.
In addition, due to
the linear duality theory,
we have $z_k=\Theta_k$.
On the other hand, since $s_k$ and $z_k$ are convex combinations of the
objective function and the constraint function in
(\ref{eq:opt_total_EED_prob}) and (\ref{eq:opt_constraint_ratio}),
respectively, we immediately arrive at
$(s_k,z_k)\in C$ and hence $z_k \ge \hat{p}^*$ by definition
of $\hat{p}^*$ for convexification of the primal problem {\bf Popt}.

For convergence, it is also observed that
$\{ z_1,\cdots,z_k \}$ and $\{ \Theta_1,\cdots,\Theta_k \}$ are
non-increasing sequences (follows from that $DLP^{k+1}$ has one more constraint than $DLP^{k}$)
and will converge at a certain optimal value at $\hat{p}^*$
($z_k$ is bounded below by zero according to the
 inequality of EED in (\ref{eq:eed_nnequality})).

Further, in Step 3) the
dual function can be readily solved as the constraint in
(\ref{eq:opt_constraint_ratio}) is relaxed, and searching the points in
the feasible domain of the non-negative $L$-dimensional vector space
$\mathbb{R}_{\ge 0}^L$ (a closed convex domain) can be readily implemented.
For the convergence property of the sequence $\{ \nu^{(i)}_k \}$,
below it will be shown that $\{ \nu^{(i)}_k \}$ is a bounded sequence.
Combining Step 3) with Step 2), by selecting an
initial point $\boldsymbol{\beta}_{1}^{(\boldsymbol{i})}$
satisfying $\mathbf{1}^T \boldsymbol{\beta}_1^{(\boldsymbol{i})} < 1$
in the initialization step,
it is observed from
(\ref{eq:generalized_programming_LP_dual_con1})
that, all elements of the dual variable sequence
$\{ \nu^{(i)}_k \}$ must satisfy the inequality as follows,
\begin{align}
&\quad \Theta_k - \sum_{i=s,r}\nu^{(i)}_k\left(\mathbf{1}^T \boldsymbol{\beta}_1^{(\boldsymbol{i})} - 1 \right)
\le \sum_{n=1}^{N} c_n \overline{EED}_{d_n}(\boldsymbol{\beta}_{1}^{(\boldsymbol{i})}), \quad \forall k \label{eq:alg_converge_1}
\end{align}
Taking into account the fact that $\nu^{(i)}_k \ge 0$ and
the assumption that $\mathbf{1}^T \boldsymbol{\beta}_1^{(\boldsymbol{i})} < 1$,
from (\ref{eq:alg_converge_1}) we have,
\begin{align}
-\nu^{(i)}_k\left(\mathbf{1}^T \boldsymbol{\beta}_1^{(\boldsymbol{i})} - 1 \right)
\le& -\sum_{i=s,r}\nu^{(i)}_k\left(\mathbf{1}^T \boldsymbol{\beta}_1^{(\boldsymbol{i})} - 1 \right)\nonumber\\
\le& \sum_{n=1}^{N} c_n \overline{EED}_{d_n}(\boldsymbol{\beta}_{1}^{(\boldsymbol{i})})-\Theta_k
\end{align}
By taking some arithmetic operations, we hence arrive at
\begin{align}
0 \le \nu^{(i)}_k \le
\frac{\sum_{n=1}^{N} c_n \overline{EED}_{d_n}(\boldsymbol{\beta}_{1}^{(\boldsymbol{i})})-\Theta_k}
{-\left(\mathbf{1}^T \boldsymbol{\beta}_1^{(\boldsymbol{i})} - 1 \right)}, \forall k \label{eq:alg_converge_2}
\end{align}
since $\mathbf{1}^T \boldsymbol{\beta}_1^{(\boldsymbol{i})} < 1$ by assumption.
Observing that (\ref{eq:alg_converge_2}) is valid for
all $\nu^{(i)}_k$ in iteration,
%On the other hand, we have $\nu^{(i)}_k \ge 0$.
it is hence concluded that the sequence of $\nu^{(i)}_k$ has
a convergent subsequence since it is bounded.

Combining the analysis above with
the duality theory in \cite{Freund04} that
$d^*=\hat{p}^*$ ($\hat{p}^*$ is the solution to the convexified primal problem),
we hence arrive at \cite{Freund04}
\begin{align}
\lim_{ k \rightarrow \infty} g^{\bf P1}(\nu^{(i)}_k)
\le d^* = \hat{p}^* \le \lim_{k \rightarrow \infty} \Theta_k
\end{align}

Geometrically, the generalized programming can be taken
as inner convexification task for the primal problem and
outer convexification for the dual problem.
While solving $d^*$, i.e., the optimal solution to the dual problem
{\bf Popt-D}, via the generalized programming algorithm,
the obtained solution to {\bf Popt} is hence
given by
$\sum_{n=1}^N EED_{d_n}
\left( \boldsymbol{\beta}_{k+1}^{(\boldsymbol{i})} \right)$
with the fulfilled stopping criterion.

Note also that we are not able to analytically discuss
how fast the generalized programming converges,
however, in the case studies, we find tens of iterations are
sufficient to guarantee the fulfillment of the stopping criterion
in Step 4). In addition, it is observed that in each iteration,
we only need to solve the linear programming in Step 2)
as well as the dual function in Step 3). Combining this with the number of iterations needed in practice,
the computation complexity of the proposed algorithm hence
is not NP-hard.

\begin{remark}
It is noted that the generalized programming is done offline and only need to be implemented once
for use,
due to the assumption that only statistical knowledge of the
channel gains is available at the transmitter. In addition, in Step 3) of
the generalized programming in each iteration, the optimal power allocation parameter
of each layer is updated by solving (\ref{eq:generalized_programming_dual}),
where the
dual variable $\nu_k^{(i)}$ employed in (\ref{eq:generalized_programming_dual}) is updated in Step 2) of the same iteration.
By repeating such a procedure till the stopping criterion is satisfied,
it is expected that the optimal
solution to the dual problem {\bf Popt-D} is obtained. During transmission,
the transmitter and the relay will therefore
transmit symbols with the obtained optimal ratio of power allocated to each layer.
\end{remark}

\begin{remark}
It is also worthy to note that, for a message of $L$ layers,
the active relay might spend more time in decoding and
encoding by employing SIC compared with that for a single
layer message. However,
as all such operations can be done with the
advanced hardware nowadays, the incurred delay
is negligible and hence is not taken into account in this work.
\end{remark}

\section{Numerical Results}\label{sec:numerical_results}
Case studies are conducted to examine
the proposed EED model and the formulated optimization problems.
We consider a square topology with the side length denoted by $d$,
with the
source node located at the origin point ($0$,$0$)
and the relay node located at the center point ($d/2$,$d/2$),
if not otherwise noted.
All destination nodes are assumed to be placed uniformly in
this square area. Moreover, we label the point ($d/2$,$d/2$)
as the reference location point
where its average receiver-side SNR from the source
is referred to as the reference receiver-side SNR in our study,
i.e., the normalized average receiver-side power for an arbitrary
link $i \rightarrow j$ in our study is
$P_i\bar{h}_{ij}/\bar{h}_{sr}=P_i\bar{h}_{ij}(\frac{d}{\sqrt{2}})^{\alpha}$
where $P_i$ is the transmit power at node $i$ and $\alpha$ is the pathloss exponent.
In addition, we assume $\rho=2$ for the Nakagami-$\rho$ channel.
%and the noise power level
%$\frac{N_0B}{2}$ ($B$ is the noise power bandwidth which is set to be $1$Hz)
%is set to be unity
Here we consider two performance metrics in the numerical part in terms of the weighted averaged EED.
One is with $c_n=1/N$, i.e., the minimization of the averaged EED of all users, and is denoted by {\bf P1} for reference.
The other takes the fairness issue into account, i.e., we cares for the user with the worst EED by setting its weight parameter to be unity and all others to be zero, and is denoted by {\bf P2} for reference.

Further, for fair comparison, in this section, six schemes are evaluated,
including (1) the proposed relay aided JSCC scheme with three
resolution levels, (2)
the direct JSCC multicast scheme with three resolution levels,
(3) relay aided JSCC scheme with two resolution levels,
(4) direct JSCC multicast scheme with two resolution levels,
(5) the mono modulation scheme (single resolution) with relay, and
(6) the mono modulation scheme without relay. In addition,
$64$-QAM, QPSK/$16$-QAM and BPSK/BPSK/$16$-QAM are considered for
mono system, two-level system and three-level system, respectively.
It is noted that all these schemes transmit the same number of bits
per source symbol for fair comparison.
%Note that here only up to three-resolution JSCC system
%is studied to reduce computational complexity.

\begin{figure}[t]
   \centering
   \includegraphics[width = 7.2cm]{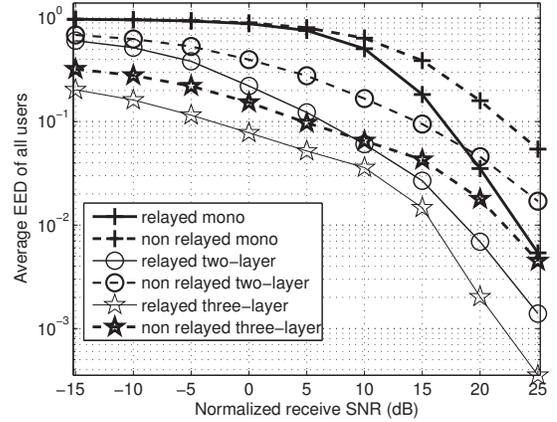}
   \caption{Comparison of the proposed relay-aided multicast approach
   with the reported counterparts in term of the average EED ({\bf P1}) under the
   same transmit power at the source and relay.} \label{fig:avg_eed}
   \end{figure}

\begin{figure}[t]
   \centering
    \includegraphics[width = 7.2cm]{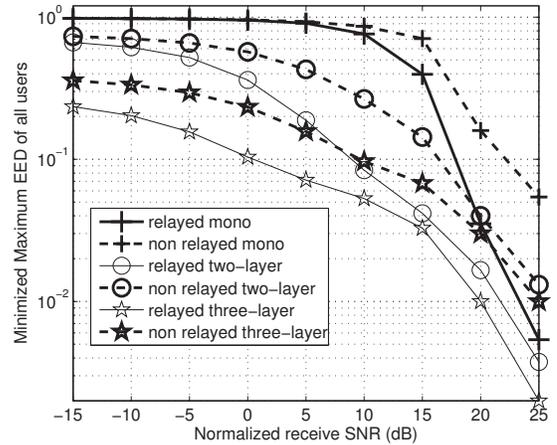}
   \caption{Minimized EED of the worst user of different schemes ({\bf P2}) under the same
   transmit power at the source and relay.} \label{fig:max_eed}
   \end{figure}

In Figs. \ref{fig:avg_eed}-\ref{fig:max_eed}, the minimized average EED of all users,
and the minimized EED of the worst user of all schemes are evaluated, respectively.
It is observed that the three-level relay aided JSCC system outperforms
all the other schemes in terms of all metrics.
In the low SNR regime, it is observed that the
three-level relay aided JSCC system greatly reduces EED
than by the mono relay aided system as well as
the two-layer relay aided system.
For instance, with normalized receiver-side
SNR $-15$dB, in the three-level relay aided JSCC system, the
averaged EED and the worst EED are less than $0.3$, as the BPSK encoded
base layer data can be possibly decoded successfully
in the extremely low SNR regime.
In the mono case, on the other hand, it
is almost impossible to decode a $64$-QAM constellation and
its EEDs of both metrics approach unity.
Even for the two-level system, the EEDs of both metrics are
higher than $0.5$, as the probability of successfully
decoding a QPSK symbol is small. It is also
observed that the gaps between different curves gradually shrink with
the increased SNR, as the
probability of successfully decoding the associated modulated symbols increases.
Interestingly with normalized SNR $15$dB, the performance of the
two-level
system is only slightly worse than that of the three-level system,
as the probability of successfully decoding a QPSK symbol is
relatively high. In other words, the base layer of
the two-level system and
the first two lower layers of the three-level system are
both decodable with a high probability,
and hence the performances of both systems are
comparable.
Meanwhile, it is not surprisingly to observe that
the mono system performs close to
the layered JSCC system with the SNR of $25$dB, i.e.,
in the high-SNR regime, for both EED metrics.

\begin{figure}[t]
   \centering
   \includegraphics[width = 7.2cm]{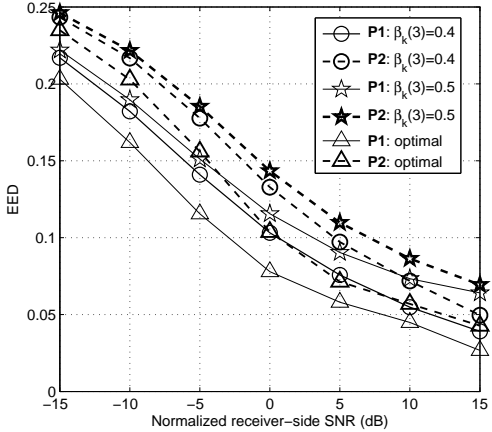}
   \caption{Performance comparison of optimized power allocation parameters
   versus randomly selected ones under the
   same transmit power at the source and relay.
   For non-optimal power vectors at the source and relay,
   $\beta_k(3)$ ($k=s,r$) is given (equal to $0.4$ or $0.5$) with
   the locally optimized $\beta_k^o(1)$ and $\beta_k^o(2)$ subject
to the constraint that $\beta_k^o(1)+\beta_k^o(2) \le 1-\beta_k(3)$.} \label{fig:parameter}
   \end{figure}

In Fig. \ref{fig:parameter}, the optimized three-level
relay aided JSCC system achieves the best performance in both of
the scenarios compared with all the other cases.
Table \ref{table:parameter} shows the optimized power allocation
vectors for the proposed relay aided three-level JSCC system.
It is observed that, with extremely low SNR ($-15$dB), almost all
power is assigned to the base layer.
With the increasing SNR,
more power assignment can be moved to higher layers
to enable high-quality reconstruction. On the other hand, it is
also observed that the optimized power allocation vectors for the averaged EED metric and the worst EED metric are slightly different, as for the latter case,
more power should be assigned to the lower layers to guarantee basic reconstruction
quality of the worst user.

%\begin{figure}[t]
%   \centering
%   \includegraphics[width = 7.2cm]{hybrid.eps}
%   \caption{Optimized Hybrid EED function of different schemes under the
%   same transmit power at the source and relay.} \label{fig:hybrid_eed}
%   \end{figure}

%\begin{figure}[!t]
%	\centering
%  \subfigure[]
%  {\label{fig:avg_eed_3d}\includegraphics[width=0.45 \textwidth]{total_distortion_3D.eps}} \hfill
%  \subfigure[]
%  {\label{fig:max_eed_3d}\includegraphics[width=0.45 \textwidth]{max_distortion_3D.eps}}
%  \caption{EED metrics of the proposed three-level relay-aided JSCC system
%  for different relay placements
%  (a) averaged EED and
%  (b) minimized EED. The normalized receive-side SNR is $5$dB.}
%\label{fig:3d}
%\end{figure}

%\begin{figure}[!t]
%   \centering
%   \includegraphics[width = 7.2cm]{large_number.eps}
%   \caption{EED performance versus the number of users.} \label{fig:large_number}
%   \end{figure}

\begin{table*} [ht]
\caption{Optimized power allocation parameters for three-level decode-and-forward (DF)
relay aided JSCC system}
\centering
\begin{tabular}{|m{2cm}|c|c|c|c|}
\hline
Normalized SNR & \multicolumn{2}{ |c| }{ Average EED }  &  \multicolumn{2}{ |c| }{Worst EED}   \\
\hline
 &  $\boldsymbol{\beta_s^*}$ &  $\boldsymbol{\beta_r^*}$ &  $\boldsymbol{\beta_s^*}$ &  $\boldsymbol{\beta_r^*}$  \\
\hline
-15dB &  $\approx(0.9,0.1)$ & $\approx(0.9,0.1)$ & $\approx(0.9,0.1)$ & $\approx(0.95,0.05)$   \\
\hline
0dB & $\approx(0.75,0.2)$  & $\approx(0.7,0.25)$ & $\approx(0.84,0.15)$ & $\approx(0.8,0.2)$ \\
\hline
15dB & $\approx(0.61,0.3)$  & $\approx(0.58,0.3)$ & $\approx(0.8,0.15)$ & $\approx(0.74,0.2)$ \\
\hline
\end{tabular}
\label{table:parameter}
\end{table*}

%Fig. \ref{fig:3d} \hl{shows that, in the averaged EED scenario, the relay
%node should be placed
%at a position close enough to the source node, so that the relay node
%can receive more layered symbols reliably from the source node
%and forward them to the destination nodes
%in the second slot to improve the averaged EED.
%In fact, it is observed that ($0.15d$, $0.15d$) is the best position for
%average EED reduction.
%However, with the worst EED metric,
%the center is the best position for placing the relay
%in order to take care of every edge user
%possibly starved due to long distances.}

%In Fig. \ref{fig:large_number}, the EED performance with $100$ destinations and
%$500$ destinations are compared,
%where for both cases destinations are
%uniformly distributed.
%It is observed that, even with $100$ destinations,
%the achieved performance is very close to that with $500$ destination
%nodes for both the averaged EED and the maximum EED metrics.
%It hence validates our discussion on signaling overhead reduction
%with a relatively
%large number of destination nodes in Sec. V.
%As observed in this case study,
%a broadcast network with $100$ users is sufficient to
%yield the sufficient accurate approximation

\section{Conclusions}
The paper studied a relay-aided joint source-channel coding (JSCC)
multicast network containing a source, a decode-and-forward (DF) relay,
and multiple receivers.
A novel EED model for a general $L$-layer scalably coded source over
fading channels was provided, which was further taken as
the performance metric for the formulated optimization problems to
determine a few key parameters such as resource allocation of
different resolution layers at the source
and relay.
A novel programming algorithm was developed to obtain a good sub-optimal
solution with guaranteed convergence. The case study results showed
that the proposed relay aided multi-resolution design yields merits
in suppressed EED against its counterparts in all the considered
scenarios. In particular, we found that with more resolutions the
EED performance could be considerably improved due to finer
granularity of quality provisioning in presence of a large number
of receivers with multi-user channel diversity.
%The paper studied a relay-aided joint source-channel coding (JSCC)
%multicast network
%containing a source, a decode-and-forward relay, and
%multiple receivers.
%EED of all source-destination pairs was evaluated and taken as
%the performance metric for the system optimization,
%to determine power allocation at the source and relay,
%and the placement of relay.
%The generalized programming algorithm was employed to
%obtain a near-optimal solution with guaranteed convergence.
%The case study results showed that the proposed relay aided
%multi-resolution design yields merits in suppressed EED against
%its counterparts in all the considered scenarios.
%In particularly, we found that with more resolutions
%the EED performance
%could be
%considerably improved due to
%finer granularity of quality provisioning in presence
%of a large number of receivers with multi-user channel diversity.

\appendices

\section{Derivation of Error Probability of Hierarchical Constellations} \label{appendix:2}
\label{appendix:2}
Here we derive the symbol error probability of each layer of
the two-layer QPSK/QPSK as an example.
It is noted that other hierarchical modulation cases
can be similarly derived, and hence here we only focus on the two-layer
QPSK/QPSK superimposed case
as shown in Fig. \ref{fig:three_layer} for brevity.
We shall firstly derive the symbol error rate of the base layer
and then decode the enhancement layer after applying
SIC for the base layer.
Assuming that $E$ is the average transmit energy for each hierarchical modulated
symbol\footnote{Note that for the average symbol energy
of each hierarchical constellation, we have $E=P/f_{sym}$ where $P$ is the transmit
power and $f_{sym}$ is the symbol rate of the hierarchical
constellation.},
we have $E_1=\beta E$ assigned to the base layer symbol and
$E_2=(1-\beta)E$ assigned to the enhancement layer symbol.
In addition, let $h$ be the instantaneous channel gain. Hence,
the received symbol energy for the base layer and the enhancement layer are
$ hE_1$ and $hE_2$
respectively.

The coordinates of each superimposed symbol can be split into the abscissa
and ordinate components, which are independently distorted by
AWGN with its two-sided power spectral density $N_0/2$.
The coordinates of the $16$ points in the constellation diagram,
$(x_i, x_j')$, are independent normal
variables with the following means and variances:
\begin{align}
&x_i \sim N(\pm \sqrt{\frac{hE_1}{2}} \pm \sqrt{\frac{hE_2}{2}},\frac{N_0}{2}) \nonumber\\
&x_j' \sim N(\pm \sqrt{\frac{hE_1}{2}} \pm \sqrt{\frac{hE_2}{2}},\frac{N_0}{2}) \nonumber
\end{align}
%where $\sqrt{\frac{E_1}{2}}$ can also be viewed as the distance between
%each base layer symbol and the origin
%and $\sqrt{\frac{E_2}{2}}$ can be viewed as the distance between
%the hierarchical modulated
%symbol and its surrounding base layer symbol.

\subsection{Symbol Error Rate of Base Layer }
For the base layer of the two-layer QPSK/QPSK,
the decision boundaries are the vertical axis for
the abscissa region
as well as the
horizontal axis for the ordinate region.
It is noted that, due to
symmetry, only symbols in the first quadrant, namely, $s_1 \rightarrow s_4$,
are considered.
For the abscissa region, the conditional base layer
error probabilities given
each transmitted SPC symbol can be
determined and categorized into two cases as follows,
\begin{align}
p_{err1|s_q}^{(1)}=
\begin{cases}
Q\left[\sqrt{\tfrac{2}{N_0}} \left( \sqrt{\frac{hE_1}{2}} - \sqrt{\tfrac{hE_2}{2}} \right) \right] & \text{if }  q = 1,3 \\
Q\left[\sqrt{\tfrac{2}{N_0}} \left( \sqrt{\frac{hE_1}{2}} + \sqrt{\tfrac{hE_2}{2}} \right) \right] & \text{if }  q = 2,4.
\end{cases}
\end{align}

Similarly for the ordinate region, the conditional base layer
error probabilities given
each transmitted SPC symbol are derived as follows,
\begin{align}
p_{err2|s_q}^{(1)}=
\begin{cases}
Q\left[\sqrt{\tfrac{2}{N_0}} \left( \sqrt{\frac{hE_1}{2}} + \sqrt{\tfrac{hE_2}{2}} \right) \right] & \text{if }  q = 1,2 \\
Q\left[\sqrt{\tfrac{2}{N_0}} \left( \sqrt{\frac{hE_1}{2}} - \sqrt{\tfrac{hE_2}{2}} \right) \right] & \text{if }  q = 3,4.
\end{cases}
\end{align}

Provided the assumption that each point is
equally likely transmitted, the
base layer error probability of the 2-layer QPSK/QPSK
given the instantaneous channel gain can then be given by,
\begin{align}
p^{(1)}=\frac{1}{4}\sum_{q=1}^4\left( 1 - \left(1-p_{err1|s_q}^{(1)}\right)\left(1-p_{err2|s_q}^{(1)}\right)  \right).
\label{eq:appendix1_layer_1}
\end{align}
It is noted that the average error probability of the base layer can be obtained by
averaging over the channel gain distribution and the details are omitted due to its simplicity.

\subsection{Symbol Error Rate of Enhancement Layer}
Note that to successfully decode the
enhancement layer, the receivers need successfully
decode the base layer and apply SIC for the base layer
before decoding the enhancement layer.
For clarity, let {\tt B} and {\tt E} denote the events where the base and enhancement
layers of one SPC symbol are correctly detected, respectively.
Applying the definition of conditional
probability, the SER of the enhancement layer ($L=2$) is expressed using the
intersection probability of events {\tt B} and {\tt E} as follows:
\begin{align}
p^{(2)}&=1-P({\tt B} \cap {\tt E})=1-P({\tt B})P({\tt E} | {\tt B}) \nonumber\\
&=1-\left(1-p^{(1)}\right)\left(1-p_{{\tt cond}}^{(2)}\right) \label{eq:appendix1_layer_2}
\end{align}
where $p^{(1)}$ is the SER of the base layer and $p_{{\tt cond}}^{(2)}$ is
the conditional SER of the enhancement layer provided
correct reception of the base layer.
Noting that SIC is applied for the base layer,
the received QPSK symbol of the enhancement layer
only has an
average energy of $hE_2$ remaining, its conditional SER hence is given by
the standard symbol error equation
for a QPSK demodulator as follows,
\begin{align}
p_{{\tt cond}}^{(2)}=2Q\left[\sqrt{\frac{hE_2}{N_0}}\right]
-Q\left[\sqrt{\frac{hE_2}{N_0}}\right]^2. \label{eq:appendix1_conditiona}
\end{align}
Incorporating (\ref{eq:appendix1_layer_1}) and (\ref{eq:appendix1_conditiona}) into (\ref{eq:appendix1_layer_2}),
the SER of the enhancement layer given the instantaneous channel gain can hence be obtained.
%Hence, they can be applied in the derivation of EED in Sec. III.
It is also noted that the average error probability of the enhancement layer can be obtained by
averaging over the channel gain distribution and the details are omitted due to its simplicity.

In addition, it is worth noting that, similar derivations of SER can be applied
to other hierarchical constellation cases and are omitted here for brevity.

%Further,
%only the first quadrant is considered due to symmetry.

%The detailed analysis is given as follows.

\section{Detailed SER Analysis of JSCC} \label{appendix:ser}
Here we present the error probability derivation, with the assumption that
the power allocated to each layer is specified and the channel realizations are known.
Conditional on the assumption that the lower $l-1$ layers have been decoded correctly,
the associated conditional symbol
error rate (SER) for the $l$th layer information over link $i$-$j$
can be computed
and is denoted by $p_{ij,{\tt cond}}^{(l)}$
($i=s,r$ and $j=r,d_1,\ldots,d_N$).
\footnote{For the interest of readers, the exact SER expression of
each layer of the two-layer QPSK/QPSK in Fig. \ref{fig:three_layer}
is derived in Appendix \ref{appendix:2}.
The derivations
for other superimposed SPC symbol cases are however omitted for brevity.}
%with $p_{ik}^{(l)} \ge 0$ ($\forall l$).
% and we have
%$$p_{ik}^{(l)}$$
Taking into account the dependence of decoding of each layer symbols,
the probability of the event that the lower $l-1$ layer symbols are
decoded successfully while the decoder fails in decoding
the $l$th layer symbol, is hence given by
$$\prod_{k=1}^{l-1}\left(1-p_{ij,{\tt cond}}^{(k)}\right)p_{ij,{\tt cond}}^{(l)}.$$
where $1-p_{ij,{\tt cond}}^{(k)}$ is the successful decoding
probability of the $k$th layer symbol conditional on the successful
decoding of the lower $k-1$ layer symbols.
In addition, it is readily observed that, when
the decoder fails in decoding the $k$th ($k<l$) layer symbol,
the higher layer
symbols (including the $l$th layer) are naturally lost.
%The associated probability is hence given by,
%$$.
Taking into account all these scenarios,
the exact SER of the $l$th information
hence is given by
\begin{align}
p_{ij}^{(l)}=&
\prod_{k=1}^{l-1}\left(1-p_{ij,{\tt cond}}^{(k)}\right)p_{ij,{\tt cond}}^{(l)} \nonumber\\
&+
\sum_{p=1}^{l-1}\prod_{k=1}^{p-1}\left(1-p_{ij,{\tt cond}}^{(k)}\right)
p_{ij,{\tt cond}}^{(p)},
\end{align}
where the first term denotes the event that the lower
$l-1$ layers are successfully decoded while
only the $l$th layer is lost in SIC decoding,
and each element of the second term in summation
denotes the event that the lower $p$th ($p<l$) layer
symbol is lost in SIC decoding
hence the higher $l$th layer is naturally lost.
Specifically, we have
$$p_{ij}^{(l)}-p_{ij}^{(l-1)}=\prod_{k=1}^{l-1}
\left(1-p_{ij,{\tt cond}}^{(k)}\right)p_{ij,{\tt cond}}^{(l)}$$
denoting the probability that only up to layer $l-1$ over link $i \rightarrow j$ is successfully decoded
given the channel realization.

Since each destination node receives JSCC symbols via both
the direct and relay links which are decoded separately,
the E2E SER of up to layer $l$ at the end of the second slot
at $d_n$,
given the
realized link gains, denoted by $p_{err,d_n}^{(l)}$,
is hence given in \ref{eq:E2E_err_pro}.
%\begin{align}
%p_{err,d_n}^{(l)}=&p_{sd_n}^{(l)}
%\left( 1- \left( 1- p_{sr}^{(l)} \right)
%\left( 1- p_{rd_n}^{(l)}  \right) \right), \label{eq:E2E_err_pro}
%\end{align}

By averaging over the distribution of all the associated channel power
gains, the expected E2E
SER of up to layer $l$ at $d_n$ is therefore given by,
\begin{align}
\bar{p}_{err,d_n}^{(l)}
=&\iiint\limits_{h_{sd_n},h_{sr},h_{rd_n}}\,
p_{err,d_n}^{(l)}(h_{sd_n},h_{sr},h_{rd_n})
f_h(h_{sd_n}) \nonumber\\
& \,\,\, f_h(h_{rd_n})f_h(h_{sr})\,
\mathrm{d} h_{sd_n}\, \mathrm{d} h_{sr}\, \mathrm{d} h_{rd_n}
\end{align}
Recall that $f_h(h_{ij})$ is the pdf of the channel power
gain over link $i \rightarrow j$.

\section{EED of $L$-Resolution Scalable Source} \label{appendix:1}
Here the EED of an $L$-resolution scalable source over a
memoryless broadcast channel with a finite size of codebook
for each layer is derived.
As shown in Fig. \ref{fig:sys}, $\boldsymbol{z}$ denotes a $M$-dimensional real-valued vector source over
the Euclidean space $\Lambda$ with its pdf $f(\boldsymbol{z})$. The associated variance per
dimension is hence determined by
$\int_{\Lambda}||\boldsymbol{z}||^2f(\boldsymbol{z}){\it d\boldsymbol{z}}/M$.
Note that here $\boldsymbol{z}$ is transmitted as an $L$-resolution scalably encoded source over a discrete
memoryless broadcast channel and characterized by a transition
matrix with its transitional probability
$Pr\{ \boldsymbol{\hat{r}} | \boldsymbol{r} \}$, where
$\boldsymbol{\hat{r}}$
is the channel output and $\boldsymbol{r}$ is the channel input.

Due to its scalably encoding nature, the Euclidean space $\Lambda$
is first partitioned into $N_1$ disjoint regions for the base layer,
denoted by $A_{k}$ ($k=1,\ldots,N_1$) where $N_1$ is the number of
codeword vectors for the base layer, i.e., the size of the
codebook for the base layer.
For the first enhancement layer, each of the $N_1$ disjoint regions
is partitioned into $N_2$ disjoint
regions, denoted by $A_{ik}$ ($i=1,\ldots,N_1$ and $k=1,\ldots,N_2$)
where
$N_2$ is the number of codeword vectors for the first enhancement layer.
Similarly, the Euclidean space can be repeatedly partitioned into more disjoint regions for the higher enhancement layers.
For the $(L-1)$th enhancement layer, each of the $\prod_{i=1}^{L-1}N_i$ regions is partitioned into $N_L$ disjoint regions, denoted by
$A_{i_1 \cdots i_L}$ ($i_l=1,\ldots,N_l$ and $l=1,\ldots,L$)
where $N_l$ is the number of codeword vectors for the $l-1$th
enhancement layer, i.e., the size of the codebook for the $l-1$th
enhancement layer.
The vector associated with the region $A_{i_1 \cdots i_L}$ is denoted by
$\boldsymbol{z}_{i_1 \cdots i_L}$. The original source is
therefore represented by the index vector $(i_1,\ldots,i_L)$ where $i_k$ represents the $k$th layer, i.e.,
the $(k-1)$th enhancement layer if $k>1$ and the base layer if $k=0$.

Let $\pi_t(i_1,\ldots,i_L)=(r_1,\ldots,r_L)$ be a one-to-one
mapping from the index vector to the channel input vector. Let
the output vector be $(\hat{r}_1,\ldots,\hat{r}_L)$ where $\hat{r}_k \in \{ r_k, e \}$
and $e$ denotes detection error. As defined in Sec \ref{sec:eed},
$\hat{p}_{err, {\tt L}_j}$ is the detection
error probability of up to layer $j$, hence is given by
\begin{align}
\hat{p}_{err, {\tt L}_j}=&\sum_{k=1}^{j-1}\hat{p}_{err, {\tt L}_k}
+Pr \{ \hat{r}_1=r_1,\ldots,\hat{r}_{j-1}=r_{j-1}, r_j=e   \},  \nonumber\\
&\quad j=1,\ldots, L.
\end{align}
Note that each channel input vector $(r_1,\ldots,r_L)$ is uniformly
distributed over $m_e=\prod_{i=1}^L N_i$, we have
\begin{align}
&\hat{p}_{err,{\tt L}_1}= \frac{1}{m_e}\sum_{\boldsymbol{r}}
Pr\{ \hat{r_1}=e|\boldsymbol{r} \},\\
&\hat{p}_{err,{\tt L}_l}=\sum_{i=1}^{l-1}\hat{p}_{err,{\tt L}_i}
+\frac{1}{m_e}\sum_{\boldsymbol{r}}
Pr\{ (r_1,\ldots,r_{l-1},\hat{r_l}=e)|\boldsymbol{r} \}, \nonumber\\
&\quad\quad\quad\quad\quad l=2,\ldots, L.
\end{align}
%where
%\begin{align}
%$A=\prod_{i=1}^L N_i$
%&B=\prod
%\end{align}
%the channel input vector $\boldsymbol{r}=(r_1,\ldots,r_L)$ and
%the channel output vector $\boldsymbol{\hat{r}}=(\hat{r}_1,\ldots,\hat{r}_L)$.
Given the $L$-layer output $\boldsymbol{\hat{r}}$, there are $L+1$
possible outputs:

$1$)$E[\boldsymbol{z}]$ if $\hat{r}_1=e$;

$l$)$z_{1 \rightarrow l-1}$
if $\hat{r}_1 \neq e,\ldots,\hat{r}_{l-1}\neq e,\hat{r}_l=e$; $\quad l=2,\ldots,L$

$L+1$)$z_{1 \rightarrow L}$ if $\boldsymbol{\hat{r}} \neq e$.
%\end{enumerate}

The associated crossover error probabilities are hence given by,
\begin{align}
&p_e^{\pi_t}(E[\boldsymbol{z}]|\boldsymbol{z})=
Pr\{r_1=e|\boldsymbol{r}\}, \\
&p_e^{\pi_t}(z_{1 \rightarrow l-1}|\boldsymbol{z})
=Pr\{r_1,\ldots,r_{l-1},r_l=e|\boldsymbol{r}\}, \nonumber\\
&\quad\quad\quad\quad\quad\quad\quad\quad l=2,\ldots,L, \\
&p_e^{\pi_t}(z_{1 \rightarrow L}|\boldsymbol{z})=
Pr\{\boldsymbol{\hat{r}} \neq e |\boldsymbol{r}\}.
\end{align}
where $E[\boldsymbol{z}]$ is proved to be optimal when error is found at the base layer in \cite{Xiang-tit}.
Note also that EED is the mean squared error
distortion between the original source $\boldsymbol{z}$ and the output
$\boldsymbol{\hat{z}} \in \{ E[\boldsymbol{z}], z_{1 \rightarrow l}   \}$ ($l=1,\ldots,L$).
Therefore, with the error probability derived above,
it is given as follows,
\begin{align}
D_{{\tt L}_L}^{\pi_t}=&\frac{1}{M}\sum_{\boldsymbol{z}}\int_{\boldsymbol{z}
\in \Lambda}|| z_{1 \rightarrow L} -  \boldsymbol{z}  ||^2
p_e^{\pi_t}(z_{1 \rightarrow L}|\boldsymbol{z})
f(\boldsymbol{z})\textit{d}\boldsymbol{z} + \cdots \nonumber \\
& +\frac{1}{M}\sum_{\boldsymbol{z}}\int_{\boldsymbol{z}
\in \Lambda}|| z_{1 \rightarrow l} -  \boldsymbol{z}  ||^2
p_e^{\pi_t}(z_{1 \rightarrow l}|\boldsymbol{z})
f(\boldsymbol{z})\textit{d}\boldsymbol{z}\nonumber \\
&+\cdots + \frac{1}{M}\sum_{\boldsymbol{z}}\int_{\boldsymbol{z}
\in \Lambda}|| \boldsymbol{z}  ||^2
p_e^{\pi_t}(E[\boldsymbol{z}]|\boldsymbol{z})
f(\boldsymbol{z})\textit{d}\boldsymbol{z} \nonumber \\
%\label{eq:appendix_1}\\
=&\left(1-\hat{p}_{err,{\tt L}_L}\right)\frac{1}{M}\sum_{\boldsymbol{z}}\int_{\boldsymbol{z}
\in \Lambda}|| z_{1 \rightarrow L} -  \boldsymbol{z}  ||^2
f(\boldsymbol{z})\textit{d}\boldsymbol{z} + \cdots \nonumber \\
&+\left(\hat{p}_{err,{\tt L}_{l+1}}-\hat{p}_{err,{\tt L}_l}\right)
\frac{1}{M}\sum_{\boldsymbol{z}}\int_{\boldsymbol{z}
\in \Lambda}|| z_{1 \rightarrow l} -  \boldsymbol{z}  ||^2
f(\boldsymbol{z})\textit{d}\boldsymbol{z}  \nonumber \\
& + \cdots +\hat{p}_{err,{\tt L}_1} \frac{1}{M}\sum_{\boldsymbol{z}}\int_{\boldsymbol{z}
\in \Lambda}||  \boldsymbol{z}  ||^2
f(\boldsymbol{z})\textit{d}\boldsymbol{z} \nonumber \\
%\label{eq:appendix_2}\\
=&D_{Q_{L}}\left(1-\hat{p}_{err,{\tt L}_L}\right)
+\sum_{l=1}^{L-1}D_{Q_{l}}\left(\hat{p}_{err,{\tt L}_{l+1}}-\hat{p}_{err,{\tt L}_l}\right)\nonumber\\
&+\sigma^2\hat{p}_{err,{\tt L}_{1}} \label{eq:appendix_3}
\end{align}
where (\ref{eq:appendix_3}) follows from the definition of
$\hat{p}_{err,{\tt L}_l}$ and $D_{Q_{l}}$
denotes the quantization distortion of the reconstruction of
the $l$-layer resolution.

%It is also noted that the derivation above still holds true
%for the codebook with infinite number of codewords
%by setting $N_l$ approaching $\infty$.

%For the users with poorer link gains, i.e., only the lower
%$l$ layers are decodable, the EED can be simplified as
%\begin{align}
%D_{{\tt L}_l}^{\pi_t}=&D_{Q_{l}}\left(1-\hat{p}_{err,{\tt L}_l}\right)
%+\sum_{1}^{l-1}D_{Q_{k}}\left(\hat{p}_{err,{\tt L}_{k+1}}
%-\hat{p}_{err,{\tt L}_k}\right)\nonumber\\
%&+\sigma^2\hat{p}_{err,{\tt L}_{1}}
%\end{align}
%as $\hat{p}_{err,{\tt L}_k}=1$ for $k>l$.
%More interestingly, the validness of the
%channel gain categorizing in Sec \ref{sec:channel_model}
%related to the SER of decoding up to an arbitrary
%layer is therefore verified.

\end{document}